\documentclass[5p,times,authoryear]{elsarticle}
\usepackage{amsmath,amssymb,amsthm,textcomp,wasysym}
\usepackage{hyperref}
\usepackage[spanish]{babel}     
\addto\captionsspanish{%
}
\usepackage[latin1]{inputenc}   
\usepackage{flushend}           
\usepackage[figuresright]{rotating}
\usepackage{subfigure}

\begin{document}

\begin{frontmatter}
\title{Dusk Over Dawn O$_2$ Asymmetry in Europa's Near-Surface Atmosphere}

\author{Apurva V. Oza $^{1, 2}$ $^*$, Francois Leblanc $^{2}$, Robert E. Johnson$^{3,4}$, Carl Schmidt $^{2,5}$, Ludivine Leclercq $^{3}$, Timothy A. Cassidy $^{6}$, Jean-Yves Chaufray $^{2}$  }
\address{$^{1}$ Physikalisches Institut, Universit\"{a}t Bern, Bern, Switzerland \\

$^{2}$ LATMOS/IPSL, UPMC Univ. Paris 06 Sorbonne Universit\'{e}s, UVSQ, CNRS, Paris, France \\

$^{3}$ Engineering Physics, University of Virginia, Charlottesville, Virginia, USA\\

$^{4}$ Physics, New York University, New York, NY, USA \\

$^{5}$ Center for Space Physics, Boston University, Boston, USA \\

$^{6}$ Laboratory for Atmospheric and Space Physics, University of Colorado, Boulder, USA \\
}
\fntext[myfootnote]{$^*$apurva.oza@space.unibe.ch}





\begin{abstract}
The evolution of Europa's water-product exosphere over its 85-hour day, based on current models, has not been shown to exhibit any diurnal asymmetries. Here we simulate Europa's exosphere using a 3-D Monte Carlo routine including, for the first time, the role of Europa's rotation on the evolution of exospheric molecules tracked throughout the orbit. In this work we focus on understanding the behavior of a single atmospheric constituent, O$_2$, sputtered by a trailing hemisphere source with a temperature-dependence under isotropic plasma conditions as also modeled by previous works. Under rotation, the O$_2$ is also subject to the centrifugal and Coriolis forces in addition to the standard gravitational forces by Jupiter and Europa in our model. We find that the O$_2$ component, while global, is not homogeneous in Europa local time. Rather, the O$_2$ consistently accumulates \textit{along the direction of Europa's rotation} at the dusk hemisphere. When rotation is explicitly \textit{excluded} in our simulations, no diurnal asymmetries exist, and any accumulation is due to the prescribed geometry of the sputtering source. We find that the assumed thermal-dependence on the O$_2$ source is critical for a diurnal asymmetry: the diurnal surface temperature profile is imprinted on to the near-surface O$_2$ atmosphere, due to small hop times for the non-adsorbing O$_2$, which then effectively \textit{rotates} with Europa. Simulation tests demonstrate that the diurnal asymmetry is \textit{not} driven by the thermal inertia of the ice, found to have only a weak dependence ($ < 7\%$). Altogether, the various test cases presented in this work conclude that the dusk-over-dawn asymmetry is driven by Europa's day-night O$_2$ cycle synchronized with Europa's orbital period based on our model assumptions on O$_2$ production and loss. This conclusion is in agreement with the recent understanding that a non-adsorbing, rotating O$_2$ source peaking at noon will naturally accumulate from dawn-to-dusk, should the O$_2$ lifetime be sufficiently long compared to the orbital period.
Lastly we compare hemispherically-averaged dusk-over-dawn ratios to the recently observed oxygen emission data by the Hubble Space Telescope. We find that while the simulations are globally consistent with brighter oxygen emission at dusk than at dawn, the orbital evolution of the asymmetries in our simulations can be improved by ameliorating the O$_2$ source \& loss rates, and possibly adsorption onto the regolith.

\end{abstract}

\begin{keyword}
\texttt{Exospheres}\sep O$_2$ \sep Atmospheres \sep Aurorae
\end{keyword}

\end{frontmatter}



\section{Introduction}
\linespread{2.0}
Europa's molecular oxygen is of keen interest as it has been suggested to be a possible source of $\sim$ 0.1- 100 kg/s of O$_2$ (\citealp{bob03} ; \citealp{hand07_asbio}; \citealp{greenberg10}) to Europa's putative saltwater ocean, thought to have a chemistry analogous to Earth's oceans \citep{vance16}. 
This range of oceanic source rates can in principle be connected to the production and dynamics of the tenuous oxygen atmosphere, or exosphere, which we simulate in this work. \\

The origin of Europa's water-product exosphere is ultimately linked to its neighboring satellite Io. Since the Voyager spacecraft first confirmed the presence of energetic ions trapped in the Jovian magnetosphere, the exosphere has been modeled to be predominantly generated by the bombardment of oxygen and sulfur ions ultimately sourced by Io's extreme volcanism \citep{peale78}.  These energetic ions both electronically excite and impart significant momentum to the water molecules in Europa's surface triggering dissociation and chemistry \citep{bob13}. The energy released by these processes leads to the ejection, or \textit{sputtering}, of water-products: primarily H$_2$O, H$_2$, O$_2$, and trace amounts of OH, H, O, and H$_2$O$_2$, which then populate the exosphere. We focus here on a single component: the molecular oxygen, to isolate our discussion to the \textit{ orbital evolution} of the neutral exosphere. In particular, we focus on the observed near-surface component (Hall et al. 1998) derived to have a variable column density between $2 - 14 \times 10^{14}$ O$_2$ \; cm$^{-2}$ at an altitude below 400 km. The origin and fate of this O2 (Johnson et al. 2018), as well as a comprehensive overview of Europa's global exosphere (Plainaki et al. 2018) has been recently reviewed. \\

\begin{figure*}[ht]
  \centering
  \includegraphics[scale=0.5]{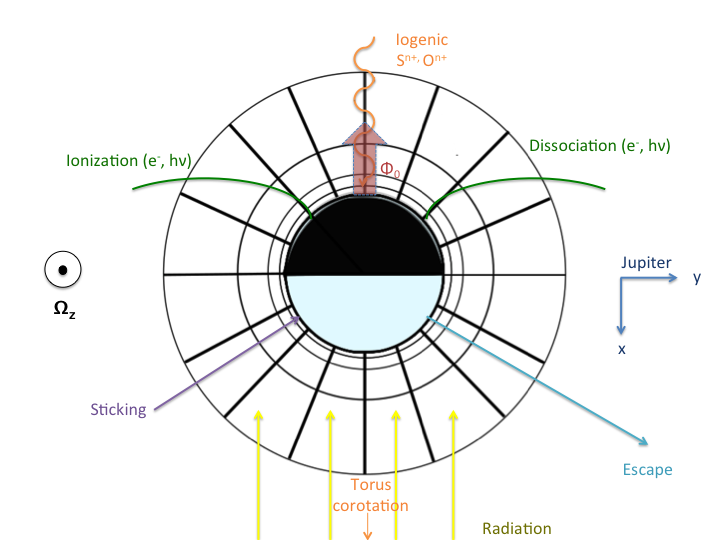}
  \caption{ Top-down geometry of the simulation domain and associated physical processes at a sub-observer longitude of $\phi_{obs} = 90$ deg, corresponding to the sunlit leading hemisphere. The O$_2$ ejection flux due to sputtering, highlighted by the red vector $\Phi_0$, dominates ejection at the sub-plasma point against the ram direction of the plasma, following Equation \ref{halfangle}. The ejected O$_2$ is assumed to be thermalized to the local surface temperature and is subject to dissociation and ionization. Adsorption, or sticking, is indicated in violet.}
  \label{grid}
\end{figure*}

In order to focus primarily on the influence of rotation on Europa's near-surface atmosphere, we employ a ballistic, 3-D Monte Carlo Exosphere Global Model (EGM: \S \ref{EGM} ). The EGM simulates Europa's neutral exosphere as it orbits Jupiter in Europa's rotating frame (EphiO). The neutral O$_2$ exosphere is governed by various physical processes illustrated in Fig. \ref{grid}. The simulations demonstrate how Europa's near-surface O$_2$ exosphere behaves in the absence of detailed knowledge of influences such as the putative H$_2$O plumes (\citet{rothwater}; \citet{roth14b}; \citet{sparkswater}; \citet{teolis16} ; \citet{sparks17}) and the near-surface plasma conditions. Although such uncertain effects could be adjusted to fit the observations, here we focus on the neutral dynamics, the effect of the source distribution, and the rotation. Our work includes the influence of Europa's rotation on the near-surface O$_2$ exosphere for the first time. The simulations are carried out in Europa's rotating reference frame, where the Coriolis and centripetal forces act on the exospheric molecules under a gravitational potential dictated by both Europa and Jupiter's gravitational fields.

\section{Europa Exosphere Global Model}\label{EGM}
Europa's water product exosphere is simulated here by tracking representative particles (H$_2$O, H$_2$, O$_2$, OH, H, and O) in a rotating, non-inertial reference frame, in spherical coordinates $(r, \theta, \phi)$, centered on Europa and extending up to $\sim 15 R_{Eu}$. We employ a parallelized Monte Carlo routine which is the core of the Exosphere Global Model (EGM). The numerical routines are similar to those described in detail in \citet{ganymede}, as well as in a companion paper on Ganymede's orbital evolution \citep{leblanc16}, where the collisionless Boltzmann equation is solved for an imposed function distribution at the surface $f_i$, for a species $i$, a velocity $v$, and the acceleration due to gravity $\textbf{g} = \frac{GM_{Eu}}{r^3} \textbf{r}$, where $\textbf{r}$ is the radial distance from Europa's center . For a more detailed description of the EGM we refer the readers to \citet{ganymede} and Section 2.2 of \citet{ludivine}. In the EGM we eject $\sim 10^{7}$ test particles from Europa's surface following known energy and angle distributions,  $f(E, \theta)$ as described later (\S \ref{sputtering}). These test particles are on ballistic (collisionless) trajectories, and can escape, stick, and be re-emitted from the surface depending on their properties as indicated in Fig. \ref{grid}. The fate of ions is not considered in this work. In the case of O$_2$, we track its interaction with the surface and account for electron and photon induced dissociation as well as ionization. Following dissociation, we also track the O formed. This procedure is carried out until a steady state O$_2$ exosphere is achieved. Steady state in our simulations is achieved when the computed thermodynamic quantities change by less than 10\% from one orbit to the next for any orbital position. These thermodynamic quantities are calculated based on the statistical weights of each test particle: $ W_i = \Phi_i \; A_{cell} \; N_i dt$ . Here $\Phi_i$ is the flux of ejected test particles, $A_{cell}$ the area of the grid-cell, $N_i$ is the number of time steps between each test particle ejection, and $dt$ is the time step between two successive positions of the test-particles, chosen so that these two positions are within one cell or two nearby cells. After a sufficient number of time-steps to achieve approximate steady state, the average number density $\langle n \rangle$, temperature $\langle T \rangle $, and velocity $\langle v \rangle $  are calculated in each cell on a spherical grid, and at seven orbital positions, $\phi_{obs}$ the sub-observer longitude of the satellite defined as $\phi_{obs} = 0 ^{\circ}$ when Europa is in eclipse behind Jupiter. In our analysis, we display the calculated densities at seven sub-observer longitudes: $\phi_{obs} = 0 ^{\circ} $, $\phi_{obs} = 20 ^{\circ} $, $\phi_{obs}= 90 ^{\circ} $ (sunlit leading), $\phi_{obs} = 150 ^{\circ} $, $\phi_{obs} = 210 ^{\circ} $, $\phi_{obs} = 270 ^{\circ} $ (sunlit trailing), and $\phi_{obs} = 340 ^{\circ} $ . To improve statistics, these quantities are reconstructed on intervals spanning six degrees in orbital longitude, around each orbital position, corresponding to a total of 1.4 hours.


\subsection{Sputtering.}\label{sputtering}
The O$_2$ exosphere is thought to be produced primarily by ions originating in the Io-plasma torus. The torus, composed predominantly of sulfur and oxygen ions trapped in Jupiter's magnetic field rotates, on average, seven times as fast as Europa, with the plasma primarily impacting Europa's trailing hemisphere in the corotation direction ($\bf{+x}$). Evidence for preferential bombardment of the trailing hemisphere by heavy ions has been observed by the Galileo Near-Infrared Mapping Spectrometer (NIMS) \citep{carlson99_so2}. Following \citep{f02}, we model the effective ion flux $\Phi_{eff}$, onto Europa's trailing hemisphere as diminishing with the corotation longitude $\phi '$:
\begin{equation}\label{halfangle}
\Phi_{eff} (\phi ') = \frac{ \Phi_{i}}{4 \pi }cos(\frac{1}{2} (\phi' - \pi)) 
\end{equation}

Where the planetary longitude is 90$^{\circ}$ out of phase with the corotation: $\phi = \phi ' + \pi/2$. 
Molecules are ejected from the trailing surface by the energy and momentum deposited by the incident ion flux $\Phi_{i}$ in the water ice regolith. Here we use sputtering rates summarized in \citet{cass13} to discuss the fate of the oxygen component. We note that plasma diversion can affect the incoming flux of all ions and electrons, and thus the exact sputtering rates are still uncertain at present \citep{cassidy16}. 
In \citet{cass13} both the hot and thermal components of the incident plasma are accounted for as well as the regolith porosity, assuming the surface is primarily pure water ice.  \citet{cass13} construct a temperature-dependent yield as the number of molecules ejected per incident ion based on \citet{fama08} and \citet{bob08}: 
\begin{equation}\label{Y}
Y(T_0)_i = Y_0 (1 + q_i \, e^{\frac{-0.06 eV }{ k T_0}})
\end{equation}

In Equation \ref{Y}, $T_0$ is the local surface temperature, Y$_0$ the yield at low temperature, and q$_i$ scales the temperature dependence of the yield for a species $i$. For the total yield given, in terms of equivalent water molecules, q$_{H2O}$= 200. The globally averaged O$_2$ ejection rate also depends on the surface temperature of the ice, with $q_{O2} = 1000$, empirically derived from extensive laboratory data described in (\citet{teolis10}; \citet{teolis17ice}), is $10^{26}$ O$_2$ s$^{-1}$, with an H$_2$ yield that is twice this rate. \citep{cass13} then estimates a globally averaged sputtering rate: $2 \cdot 10^{27}$ H$_2$O s$^{-1}$, of which the primary ejecta are H$_2$O, H$_2$, O$_2$ with trace amounts of O, H, OH.

The energy distribution, $f(E)$, of the ejected particles is modeled according to the species in question. The H$_2$O, OH, O, and H  are assumed to follow a water-ice sputtering distribution, based on the heavy ion ($E \sim keV$) sputtering of H$_2$O ice \citep{bob1990}:

\begin{equation}\label{fsputt}
f(E)_{Sputt} \sim \frac{EU^x}{(E+U)^{2+x}}
\end{equation}
where x = 0.7, and the effective surface binding energy U = 0.052 eV. The highly volatile molecular species  O$_2$ and H$_2$ are produced by radiolysis. As the energetic ions penetrate the ice, they break bonds leading to chemistry and the temperature dependent formation of H$_2$ and O$_2$. The initial ejection energy of O$_2$ (and H$_2$) can also be described by Equation \ref{fsputt} using $x=0$ and U = 0.015 eV  \citep{bob1990}. We refer to this distribution as $f(E)_{U2}$ in this work. After the first ejection, those molecules returning to the surface can either react or are subsequently thermally desorbed so that their ejection speeds are roughly determined by a Maxwell Boltzmann (MB) distribution at the surface. The Maxwell Boltzmann speed distribution, accurate for a localized point is given by: 

\begin{equation}\label{fMB}
f_{MB}(v) dv = 4\pi \left ( \frac{m}{2\pi k_b T_0} \right ) ^{3/2} v^2 exp \left (\frac{-mv^2}{2 k_b T_0} \right ) d v
\end{equation}
This distribution applies since the desorbed molecules undergo multiple interactions with the surface, which we model to be fast relative to the production of a new O$_2$, and are thus in local thermodynamic equilibrium with the surface ice, with an average energy, $\frac{3}{2} k_b T_0$. \\
	
\subsection{Surface Temperature}\label{insolation}
The effective temperature of the surface ice, $T_0$ is defined by radiative equilibrium. In our model, we consider a slab of ice with conductivity $\kappa_T$ = 2.52 $\times$ 10$^2$ erg\,cm$^{-1}$\,K$^{-1}$\,s$^{-1}$  \citep{abramovspencer08}, and calculate the equilibrium temperature across the surface, over latitude $\theta$ and planetary longitude $\phi$:

\begin{equation}
\label{eqn:Temperature_model_thermal_inertia}
\frac{F_{\astrosun, 1AU}\left(1-A\right)}{R^2 }cos\theta cos\phi = \varepsilon \sigma T_0^4 - \kappa_T \frac{\partial T}{\partial r}
\end{equation}

\noindent $F_{\astrosun, 1AU}$ is the total solar irradiance in erg\,s$^{-1}$\,cm$^{-2}$ at 1 AU, $R$ the orbital distance in AU, and $\sigma$ the Stefan-Boltzmann constant. The albedo, $A$, is assumed to vary linearly with $\phi$ between 0.65 on the leading hemisphere and 0.45 on the trailing, approximating the most recent interpretation of Galileo PPR results \citep{rath14}. The emissivity of the ice, $\varepsilon$, is set at 0.96. A small correction for the latent heat of sublimation is included \citep{abramovspencer08}. Endogenic heatflow, as estimated by \citet{spencer99} with an upper limit of 5 $\times$ 10$^8$ erg\,s$^{-1}$\,cm$^{-2}$, is neglected in this work.

\par The temperature gradient at the surface, $\frac{\partial T}{\partial r}$ in the last term of Eqn. \ref{eqn:Temperature_model_thermal_inertia} is given by the one dimensional equation for heat conduction along a direction $r$ that is \textit{normal} to the surface: 

\begin{equation}
\label{eqn:Temperature_thermal_inertia}
\rho C_P  \frac{\partial T}{\partial t} = \frac{\partial}{\partial r} \left( \kappa_T \frac{\partial T}{\partial r}\right)
\end{equation}

\noindent where $\rho$ is the mass density, taken as 0.92 g\,cm$^{-3}$, and the specific heat at constant pressure, $C_p$, is 1.96 $\times$ 10$^7$ erg\,g$^{-1}$\,K$^{-1}$ \citep{VanceGoodman09}. Eq. \ref{eqn:Temperature_thermal_inertia} is solved with the aid of a \href{http://www.boulder.swri.edu/~spencer/thermprojrs}{thermal routine} using the slab boundary method (\citet{spencer89}; \citet{young17}). Forcing of solar insolation over a synodic day in late Feb. 2015 is determined in 2$^{\circ}$ increments of $\theta$ and $\phi$ using the \texttt{309.bsp kernel} provided by JPL. While this approach simplifies the variation of thermophysical parameters over the surface \citep{rathbun2010}, it provides a basis for longitudinal asymmetries about the subsolar axis that result from the thermal inertia of the ice. 

\begin{figure*}\label{dtdphi0}
\centering
\subfigure[a]{\label{dtdphi}\includegraphics[height = 7.5cm, width=75 mm]{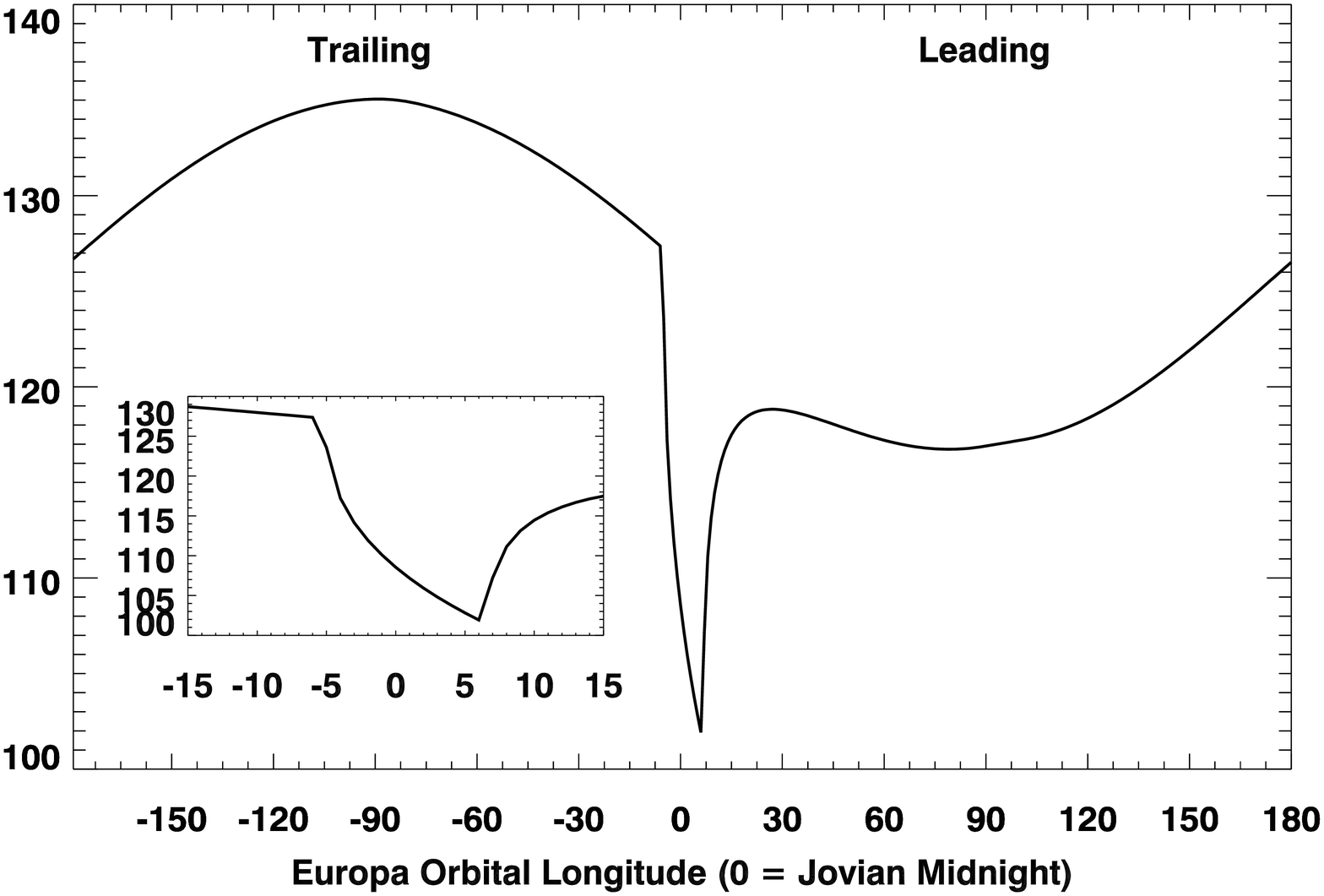}}
\subfigure[b] {\label{Tmap}\includegraphics[height = 75mm, width=85 mm]{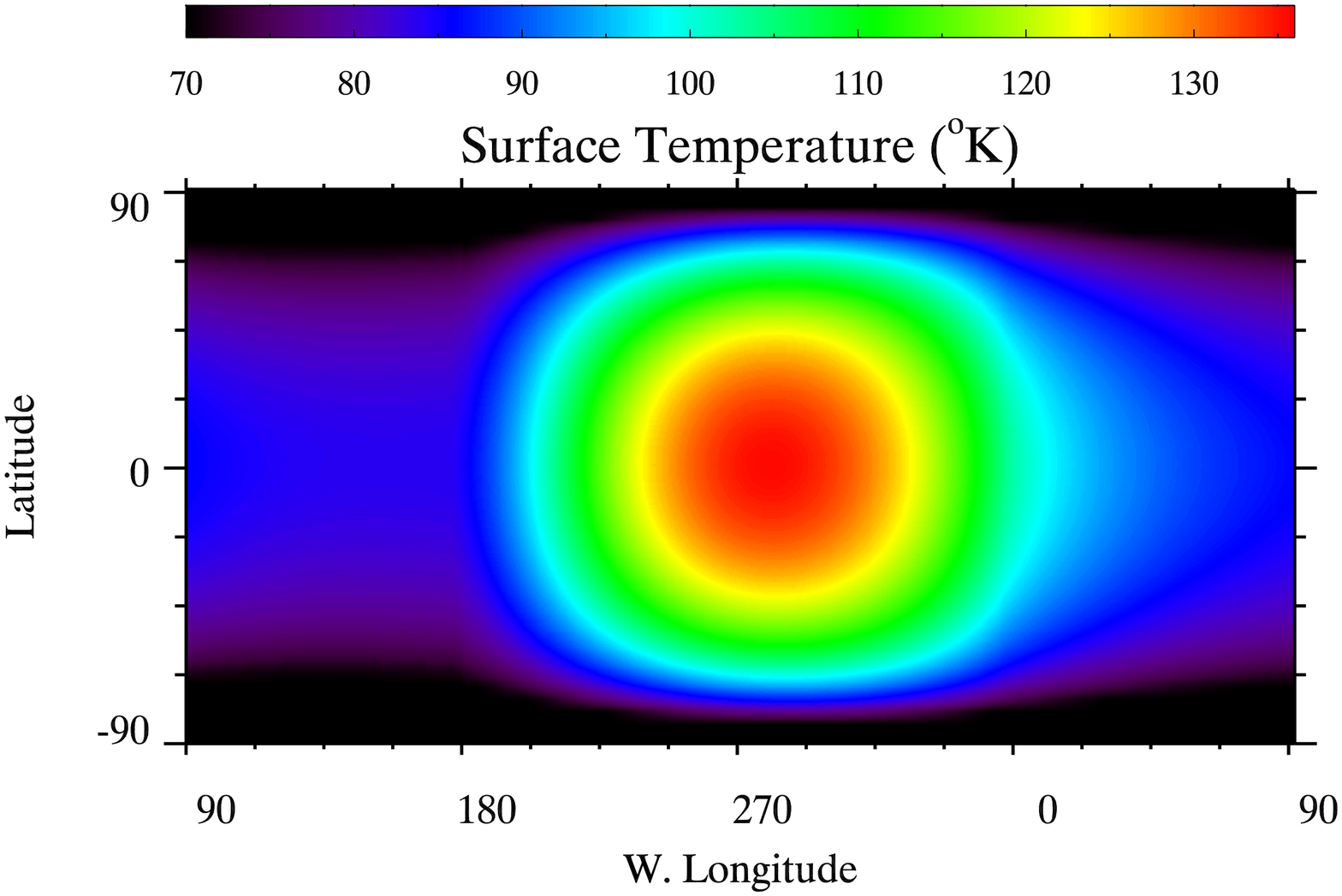}}
\caption{\textbf{a}: Equatorial subsolar surface temperature (K) as a function of sub-observer longitude, or Europa's orbital longitude, demonstrating the leading-trailing temperature profile asymmetry due to albedo. The darker, trailing hemisphere (A$_b$ = 0.45) is seen to be $\sim 18 $ K warmer than the brighter, leading hemisphere (A$_b$ = 0.65). \textbf{b}: Surface temperature map of Europa's sunlit trailing hemisphere at $\phi_{obs} = 270 ^{\circ}$. The surface temperature rapidly increases from dawn, $\phi_{obs} = 90$ until the subsolar point at $\phi_{obs} = 0$. The surface then begins to cool slightly more slowly, passing through dusk at 18h Europa local time. The largest thermal gradients driving migration are therefore between 6h - 12h, and 12h-18h. Due to the thermal inertia of the ice, the dusk hemisphere is warmer than the dawn hemisphere. }
\end{figure*}

In Figure \ref{dtdphi} we show the evolution of Europa's equatorial surface ice temperature along its orbit in panel A, where we include the effect of Jupiter's shadow. The Figure \ref{dtdphi} inlet is a zoom of the eclipse region where Europa's temperature drops by $\sim 30$ K before rising to $\sim 110 K$ at the brighter, leading hemisphere. Figure \ref{Tmap} is the corresponding temperature map of the darker, trailing hemisphere mapped in west longitude. As is true for most planetary bodies, on average, the largest thermal gradient is between noon and midnight, or day and night, as indicated in Fig. \ref{Tmap} with a slight increase in surface temperature, $\Delta T \sim 10$ K , at dusk over dawn due to the thermal inertia of the ice. The surface temperature inhomogeneities due to thermal inertia, as we will explore below, contribute only $\sim 10\%$ to the observed dusk-over-dawn asymmetry. 



\section{Results}
Here we present two results of our 3-D simulations of Europa's water product exosphere. One in which Europa was held static, and one where Europa was able to rotate about Jupiter's axis. In this work we focus on understanding the behavior of the primary atmospheric constituent: the \textit{near surface O$_2$}.  

\subsection{Atmospheric Bulges}
\begin{figure*}[ht]
  \centering
  \includegraphics[scale=0.5]{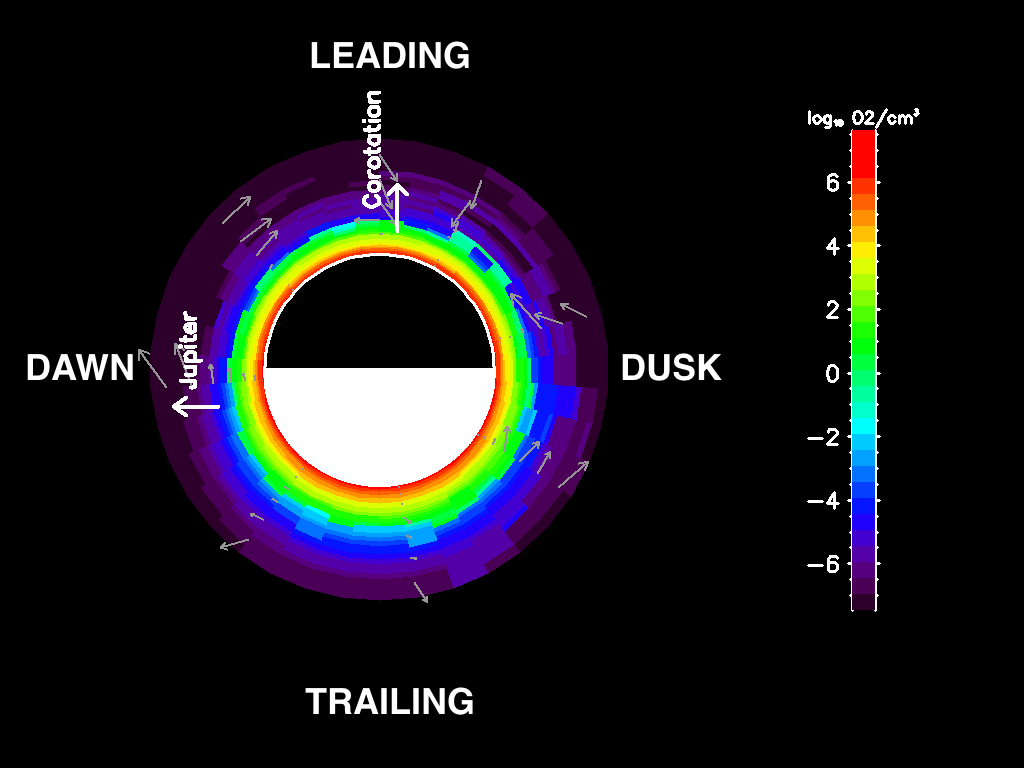}
  \caption{ Europa's static, O$_2$ exosphere for a simulation fixed at the sunlit trailing hemisphere with no rotation. Vectors are shown to demonstrate the day-night migration in the absence of rotation. An artificial day-side O$_2$ asymmetry is highlighted, due to preferential ion bombardment at the sunlit trailing hemisphere in the absence of rotation. The low albedo, sunlit trailing hemisphere has an equilibrium temperature of $T_d \sim 135 K$, further exaggerating the effect of O$_2$ production on the exosphere due to the strong thermal dependence of the O$_2$ sputtering yield. }
  \label{norot}
\end{figure*}

\subsubsection{No Rotation: Atmospheric \texorpdfstring{O$_2$ \;} BBulge at Noon} \label{diffusion}
A trailing hemisphere O$_2$ source at Europa may be expected to have a large 'atmospheric bulge' at plasma ram, where the near-surface column density is larger than the background exosphere. As O$_2$ is thought to weakly interact with the water ice, the O$_2$ migrates stochastically due to the day-night temperature gradient. The gray vectors in our simulation depicted in Figure 3, demonstrate this migration. If migration were rapid, a night-side bulge would be expected like the Helium bulges on the Moon and Mercury \citep{mercmoon}. However, migration is slow compared to the O$_2$ lifetime (c.f. Table 1) such that no asymmetry other than sputtering remains.

In a collisionless exosphere, ejected molecules return to the surface where they stick or are re-emitted with an energy determined by the local, surface ice temperature. On interacting with the surface, the fate of a molecule is sensitive to the surface residence time: $\tau_{res} = \tau_0 \; exp(\frac{E_{ads}}{k_b T_0})$ where $E_{ads}$ is the effective heat of adsorption. Initially for molecular oxygen, we assume the heat of adsorption is negligible and the surface residence time is extremely short, on the order of the vibration time for a Van der Waals potential ($\tau_0 \sim 10^{-13} s$) so that O$_2$ has a negligible sticking coefficient. The effect of O$_2$ sticking due to a longer residence time will be examined in section \ref{aurora}. Thus one can treat O$_2$ motion as a series of random walks. \citet{hunten88} for example,  defines $\lambda_0 = \frac{U_0}{k_b T_0}$, the ratio of gravitational binding energy, $U_0 = \frac{GM_p m_i}{r_p}$ to the thermal energy $k_b T_0$ at the surface, approximately equal to the ratio of planetary radius $r_p$ to the scale height, $H(T_0)$, so that $\lambda_0 \sim \frac{r_p}{H}$. $\lambda_0$, is roughly equivalent to the number of ballistic hops a molecule experiences in traveling an average distance $r_p$ yielding a diffusion timescale: $t_d \sim r_p^2 / D$, where $D$ is the effective diffusion coefficient. As approximated in \citet{hunten88}, $D \sim \frac{H^2}{t_{hop}}$, so that the migration time for traveling a distance r$_p$ at the surface is:
\begin{equation}\label{td}
t_d \sim \lambda^2 t_{hop} 
\end{equation}

\begin{table*}[t]
    \centering
    \begin{tabular}{l*{6}{c}r}
 	\bf    &  & Timescale ($\tau_{orb}$)       \\
    \hline
     Orbital Period &$\tau_{orb}$ & 1.0 & \\ 
     \emph{Ballistic O$_2$ hop} &$t_{hop}$  & \emph{$8 \cdot 10^{-4}$ }    \\  
     Day-night migration& $t_d$ & 4.2 & \\ 
     O$_2$ lifetime$^{*}$ &$\tau_{O2}$ &  1.08 & \\ 
    \end{tabular}\par
    \caption{Exospheric timescales for Europa. Given $\tau_{orb} = 3 \cdot 10^5$ sec. The average hop time for an O$_2$ molecule is computed using an average surface temperature of $\langle T \rangle \sim 110 K$, as $\sqrt{2} \frac{v_0}{g}$. The ballistic hop ,$t_{hop}$, assumes no interaction with the surface. The day-night migration time is estimated as a random-walk process across one Europa radius assuming no O$_2$ trapping or reactions with the surface ice. The O$_2$ lifetime,$\tau_{O2}$, is limited by dissociation and ionization, computed by the product of the net reaction rate coefficient and the electron number density: $(\kappa \times n_e)^{-1}$, where  $\kappa = 4.4 \cdot 10^{-8}$ cm$^3$ s$^{-1}$ is the orbit-integrated reaction rate coefficient including e- impact as well as ionization, and $n_e = 70$ cm$^{-3}$ the electron number density. }
    \label{timescales}
\end{table*}

In Table \ref{timescales}, we estimate several timescales scaled to Europa's orbital period $\tau_{orb}$, critical to the evolution of the near-surface O$_2$ in the static and rotating cases (3.1.2). Under a uniform gravitational potential, the ballistic hop time for an O$_2$ molecule is on the order of a few minutes, far shorter than the O$_2$ lifetime. Table \ref{timescales} also indicates that the O$_2$ lifetime, $\tau_{O2}$, limited by the O$_2$ production and loss rates is comparable to the orbital period, yielding an average exospheric column of $\langle N \rangle_{O_2} \, \sim 10^{14} $O$_2$ cm$^{-2}$. Thus the exosphere, based on our assumptions, should be actively built and destroyed within one orbital timescale. Lastly, it is seen that the timescale for the day-to-night flow $t_d$, is not only about four times longer than the O$_2$ lifetime in the static case, but also four times as long as the orbital period, in the rotating case. In this way, the net spreading of the O$_2$ source is on average \textit{slower} than Europa's orbital speed so we can, to first order, consider the O$_2$ as coupled to the rotating surface. We will demonstrate that this coupling is the principal mechanism for the time-dependent atmospheric O$_2$ bulge.


\subsubsection{Rotating Case: Atmospheric \texorpdfstring{O$_{2}$ \;} BBulge at Dusk.}\label{rot}

\begin{figure*}[ht]
  \centering 
  \hspace*{-1cm}
  \includegraphics[width=200 mm, height = 132 mm]{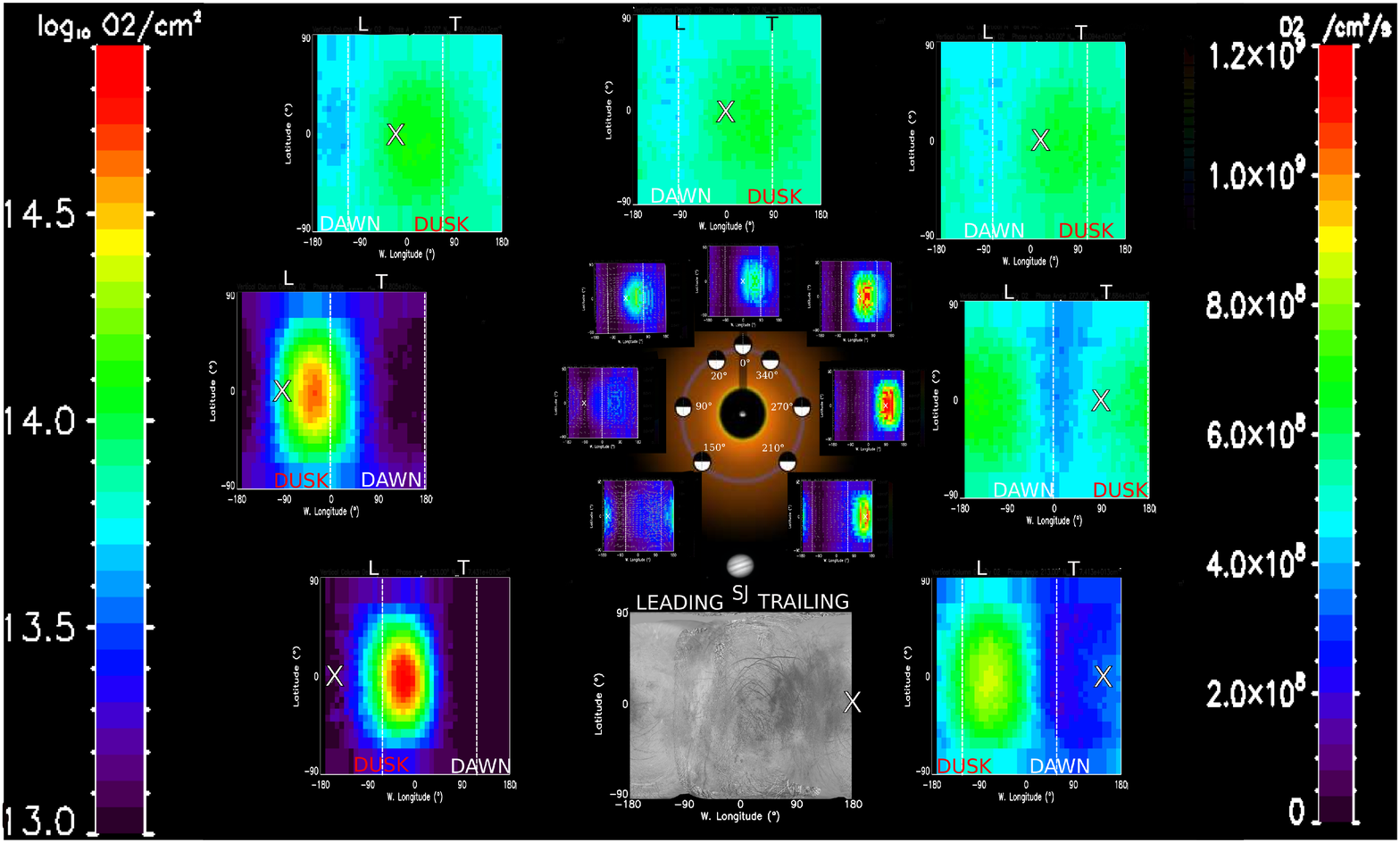}
  \caption{ Formation map of Europa's O$_2$ exosphere, illustrating surface-exosphere coupling of the near-surface O$_2$. $\phi = 0$ degrees west longitude corresponds to the subjovian point (SJ), where positive longitudes correspond to the trailing hemisphere (T) and negative longitudes, leading (L). At $\phi_{obs} = 180^{\circ}$ we provide a high-resolution map of Europa's surface, for comparison to the near-surface exosphere, taken by Voyager and Galileo as adopted in \citep{doggett09}. On each simulation map, the subsolar point is indicated by a white X, whereas the dawn and dusk terminators are indicated by the vertical dashed lines. \textbf{Inner circle surface maps}: Exospheric production as shown by the ejection rate of the O$_2$ sputtered from the surface with respect to latitude and west longitude at different orbital positions of Europa. The sputtering flux ranges from 2 (blue) - 12 (red) $\cdot 10^{8}$ O$_2$ cm$^{-2}$ s$^{-1} $ (right panel colorbar). Sub-observer longitudes($\phi_{obs}$), or phase angles are provided inside the orbit.  \textbf{Outer circle exosphere maps}: Log$_{10}$ of the radially integrated column density of the O$_2$ exosphere with respect to latitude and west longitude at different orbital positions of Europa. The radial column density ranges from $\sim 1 - 30 \cdot 10^{13}$ \; O$_2$ cm$^{-2}$ (left panel colorbar). Whereas the average radially integrated column density of the O$_2$ exosphere ranges from $7.4 - 8.1 \cdot 10^{13}$ \; O$_2$ cm$^{-2}$. The reader is encouraged to refer to the web-version of this article for a more enhanced view of the smaller plots.}
  \label{maps}
\end{figure*}

Once rotation is activated and the O$_2$ column density is tracked throughout the orbit, we simulate an atmospheric bulge  persisting throughout Europa's orbit. What is fascinating about this bulge, representing more than 10\% of the bulk O$_2$ column density, is it exhibits a \textit{diurnal dependence} peaking consistently at dusk. A diurnal dependence is of course linked to the solar cycle, which in our model only influences the surface ice temperature (Sec. 2.1). As described in Sec. 2.2, the O$_2$ production is strongly dependent on the ice temperature. In this way, one can gain an intuitive understanding of an atmospheric bulge at dusk, by reexamining the timescales in Table \ref{timescales} alone. 
Given that $t_{hop} \ll \tau_{orb}$ the O$_2$ is effectively coupled to the rotating surface and the near-surface O$_2$ can be expected to follow the diurnal temperature variations in Fig. \ref{Tmap}, and one can simply consider the atmospheric lifetime of an O$_2$ bulge over one Europa day. For instance, the nominal atmospheric bulge at the sunlit trailing hemisphere noon (Fig. \ref{norot}), will be able to survive for at least $\tau_{orb}/4$ in the direction of rotation towards the dusk hemisphere. This 90$^{\circ}$ shift from noon was derived analytically for a tidally-locked satellite atmosphere in \citet{Oza18A}. It was shown that the balance between the satellite's atmospheric lifetime and orbital period over the orbit can be expressed as a parameter $\beta = 2 \pi \frac{\tau_{O2}}{\tau_{orb}}$. The rotational shift of an atmospheric column peaking at noon for a thermally-desorbed collisionless exosphere then, can be estimated as: $\Delta \phi \sim$ tan$^{-1}(\beta)$. Based on the modeled production and loss timescales, for Europa \textit{and} for Ganymede (Leblanc et al. 2017), the O$_2$ lifetime approaches one orbital period, so that $\beta \sim 2 \pi$, yielding $\Delta \phi \sim \frac{\pi}{2}$. The 90$^{\circ}$ longitudinal shift is responsible for a maximum column at dusk and minimum column at dawn, yielding the dusk-over-dawn atmospheric asymmetry. A dusk-over-dawn asymmetry is not expected if $\beta \gg 1$ due to far slower O$_2$ loss for example, at Rhea and Dione \citet{teolis16}. On the other hand, if $\beta \ll 1$, the O$_2$ bulge would be lost too quickly, before it has time to accumulate towards dusk. \\ 

We illustrate this apparent \textit{surface-exosphere} synergy occurring throughout the orbit by using two sets of simulated 2-D spatial morphology maps in Figure \ref{maps}. The center image indicates the orbital longitudes of the seven simulated orbital positions. The inner and outer maps represent the orbital synergy between the O$_2$ surface and O$_2$ exosphere varying with rotation. The inner circle, maps the O$_2$ source flux at the surface: radiolytically sputtered ejecta in  cm$^{-2}$  s$^{-1}$ , and the outer circle maps the O$_2$ exosphere: vertical column density in cm$^2$ . On the outer maps, consistent with the inner maps, we have indicated dusk, dawn, and the subsolar position (X), along with the plasma leading (L) and trailing (T) hemispheres. The outer maps indicate O$_2$ bulges consistently accumulating near the dusk terminator. The inner maps illustrate that the O$_2$ source is also concentrated as bulges, suggesting that the atmospheric morphology is shaped by the surface ejecta.


\subsection{Orbital Evolution of Europa's \texorpdfstring{O$_2$}: Europa's Diurnal Cycle}

\begin{figure*}
\centering
\subfigure[] 
{\label{evolution}\includegraphics[trim=0 -4cm 0 1cm, height = 65mm, width=75 mm]{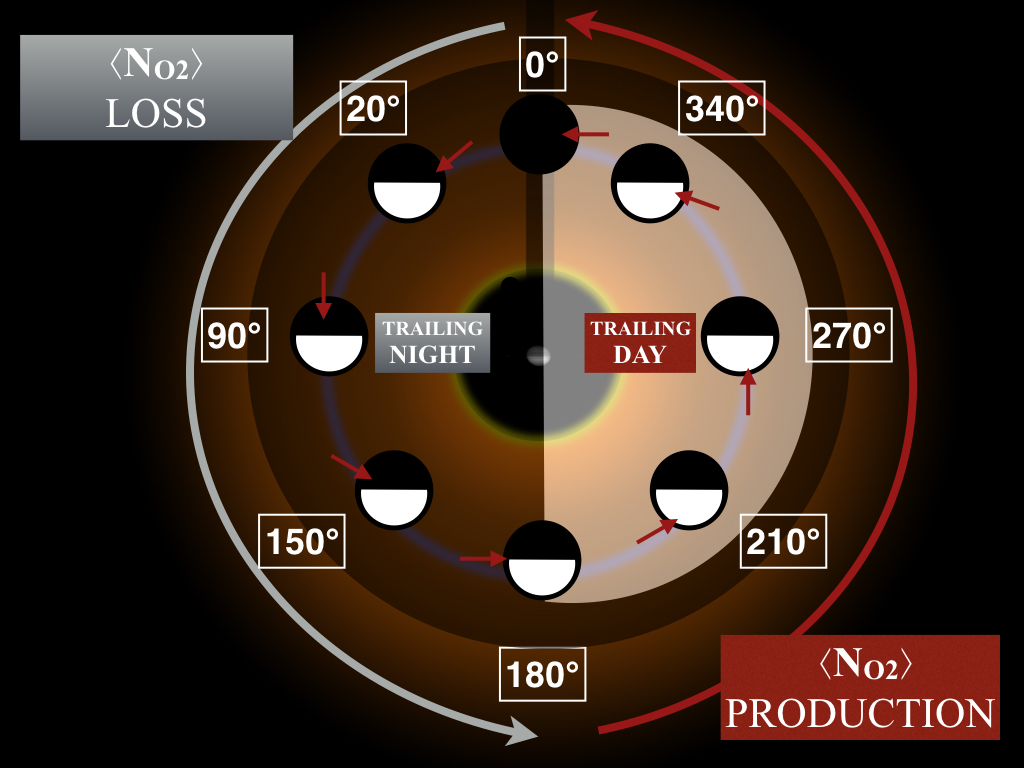}}
\subfigure[]
{\label{evol1}\includegraphics[height = 75mm, width=85 mm]{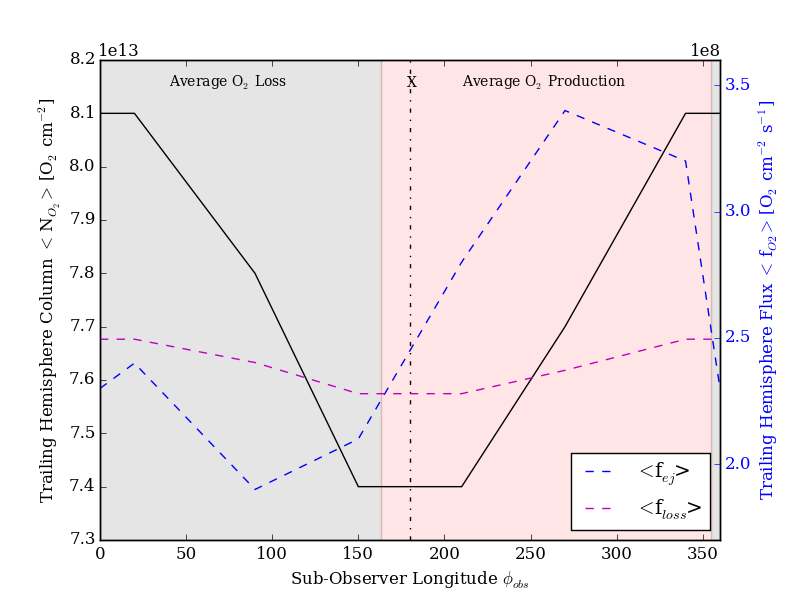}}
\caption{\textbf{a:} Illustration of Europa's diurnal O$_2$ cycle for our modeled thermally-dependent trailing hemisphere source. Day-night phases are superimposed in the frame of an observer at the trailing hemisphere. During 'Trailing Day', the exosphere grows on average ($\langle$ N$_{O_2} \rangle$ production phase). During 'Trailing Night', the O$_2$ exosphere slowly diminishes on average ($\langle$ N$_{O_2} \rangle$  loss phase). The production and loss phases corresponding to the diurnal asymmetry therefore results in the orbital evolution of Europa's O$_2$ throughout its 85-hour day.
\textbf{b:} Monte Carlo simulation results for Europa's trailing hemisphere while rotation about Jupiter is active. Orbital evolution of the surface-averaged O$_2$ column density (black) along with the surface-averaged O$_2$ production flux ( $\langle$ f$_{ej}\rangle$: blue) as well as the average O$_2$ loss flux ($\langle$ f$_{loss}\rangle$: magenta) versus the sub-observer longitude. $\langle f_{loss}\rangle$ estimates the loss of the average O$_2$ content, throughout the exospheric lifetime of the O$_2$, at each sub-observer longitude: $\langle $N$_{O_2}  \rangle  \sim \langle$ f$_{loss} \rangle \langle t_{exo} \rangle$. X indicates the day-night terminator separating the average production and loss phases. } 
\end{figure*}

Ultimately, since our modeled trailing hemisphere O$_2$ source exhibits a thermal dependence, the surface-exosphere synergies can be examined qualitatively based on Europa's day \& night phases illustrated in Figure \ref{evolution} (red vector: trailing hemisphere). Our Monte Carlo simulations in Figure \ref{evol1} show that the day \& night phases for the trailing hemisphere correspond to two different orbital phases: a globally increasing O$_2$ exosphere during the "Trailing Day" (between 180 and 360$^{\circ}$ longitude, shaded in red) and a globally diminishing O$_2$ exosphere during the "Trailing Night" (shaded in gray). Altogether, Figures \ref{evolution} \& \ref{evol1}  illustrate Europa's diurnal O$_2$ cycle, for a thermally-dependent, trailing hemisphere O$_2$ source.  Unlike the static, non-rotating case simulated in Section \ref{diffusion} and previous works, this diurnal cycle is found to peak at dusk due to atmospheric O$_2$ accumulation over time. \\

Second order effects, such as O$_2$ migration or thermal effects during eclipse, treated in our simulations, may contribute to angular deviations from a 90$^{\circ}$ longitudinal offset derived analytically for a highly thermal-dependent O$_2$ source. In our simulations, the O$_2$ bulges are vastly more apparent post-eclipse, near the sunlit leading hemisphere phases (i.e. 90$^{\circ}$  \& 150$^{\circ}$) during which the bulk O$_2$ is being lost as described in figures \ref{evol1} \& \ref{evolution}. \\

Overall, the 2-D maps of Fig. \ref{maps} reveal that the \textit{local} O$_2$ column densities vary substantially, by more than an order of magnitude $\sim$ 90\%, due to the thermally-dependent O$_2$ source fluxes also responsible for the shape of the atmospheric dusk bulges. These simulated maps provide a first attempt at a detailed picture of how Europa's rotating O$_2$ exosphere could be evolving throughout its orbit in latitude and longitude, due to the diurnal O$_2$ production likely responsible for the observed dusk-over-dawn oxygen asymmetries.
\clearpage

\section{Discussion}\label{aurora}

Observations of Europa's O$_2$ exosphere over the decades have been limited, in that the only evidence of gas phase O$_2$ has been based on ratios of the far-ultraviolet atomic oxygen emission lines first observed by the Hubble Space Telescope (HST) as aurorae (\citealp{h95}; \citealp{h98}). The ratio of the auroral line intensities $\frac{I_{1356}}{I_{1304}} \sim 2$ is suggestive of e$^{-}$ impact dissociation of O$_2$:  $O_2 + e^{-} \rightarrow O^{*} + O + e^{-} + \Delta E$. The line-of-sight O$_2$ column density then can be derived based on the electron density $n_e$ [cm$^{-3}$]and electron-impact dissociation reaction rates $\kappa $ [cm$^3$/s] (e.g. Table 2 \citet{ganymede}) from: $I_{\lambda} =  \left ( n_{e}\kappa_{O_2 + e-} \right ) N_{O_2} + \left (n_{e}\kappa_{O + e-} \right ) N_{O} $. Remarkably, the first set of HST observations confirmed the predicted O$_2$ column density within a factor of two, based on sputtering experiments and atmospheric escape estimates (\citealp{brown82}; \citealp{bob82}). Based on a re-analysis of Voyager, Galileo, and Cassini data \citet{bagenal15} demonstrated that one can expect the O$_2$ to vary by at least a factor of two. Nevertheless, our knowledge on the near-surface plasma conditions influencing the derived column densities is still severely limited---a point recently reviewed in detail by \citep{plainaki18} along with previous models of Europa's asymmetric O$_2$ exosphere (\citealt{cass07}; \citealt{plain13}). Indeed, \citealt{plain13} did reproduce asymmetries in the exosphere by modeling the uncertain temperature-dependence of O$_2$, however without rotation the model did not reproduce consistent dusk-over-dawn asymmetries.\\

Despite the uncertainties \citet{roth16} (hereafter R16), equipped with the upgraded Space Telescope Imaging Spectrograph (STIS) with a spatial resolution of 71-95 km, altered the nominal paradigm of a globally-uniform exosphere. The STIS observations were able to distinguish a two-component atmosphere in \textit{altitude} $z$ and oxygen mixing ratios: (1) \textit{near-surface O$_2$ exosphere} ($z \lesssim 400$km; $\approx$ 95\%-99\% O$_2$) and (2) \textit{corona} ($z \gtrsim 400$ km; $\lesssim$ 85 \% O$_2 $). Previous observations did however provide evidence of asymmetric oxygen emission at various longitudes, $\phi = 200 - 250^{\circ}$ (\citet{mcgrath04,mcgrath09}) and $\phi \sim 90^{\circ}$ \citep{saur11} influenced by various external mechanisms such as the plasma. R16 assessed the influence of Jupiter's plasma torus on the oxygen aurorae, and found that the \textit{polar} aurorae are periodically more intense depending on the magnetic tilt. Monitoring Europa's oxygen aurorae at such high spatial resolutions throughout the orbit additionally discerned a \textit{longitudinal}, \textit{diurnal} asymmetry of unknown origin. The dusk hemisphere was observed to be consistently brighter on average than the dawn hemisphere by a factor of $\sim$ two.  \\

\begin{figure*}[ht]
  \centering
  \includegraphics[scale=0.85]{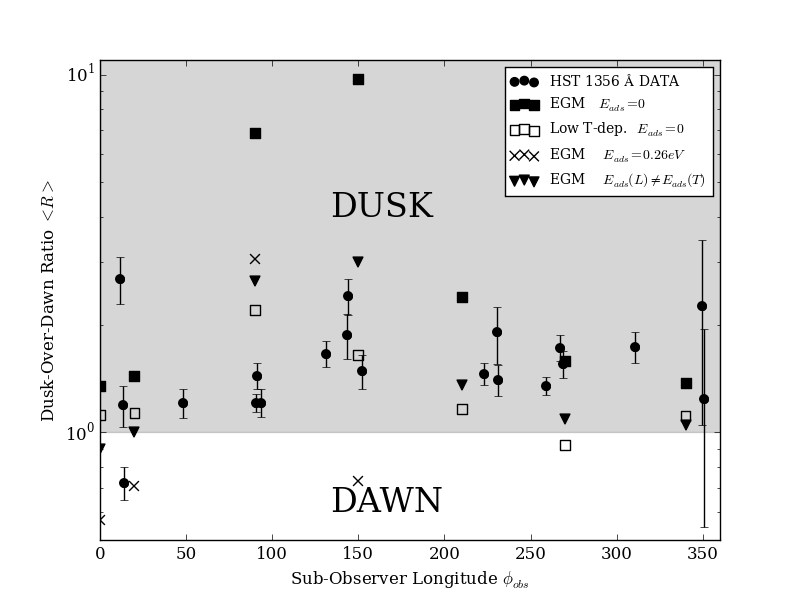}
  \caption{ Orbital evolution of the dusk-over-dawn ratio, $<R>$. Black circles with error bars: HST \textit{near-surface} 1356 $\mbox{\AA}$ auroral emission data. The average oxygen emission from the dusk hemisphere is divided by the average dawn hemisphere emission. EGM simulations of $<R>$ for an exosphere subject to isotropic plasma conditions, $n_e$ = 70 cm$^{-3}$ and $T_e$ = 20 eV, are shown for non-adsorbing O$_2$ (black squares), non-adsorbing with a small thermal dependence (open squares), highly adsorbing O$_2$ "sticking" to the surface (Xs), and a non-uniform adsorption (triangles). }
  \label{ddaurora}
\end{figure*}

\begin{figure*}
\centering
\subfigure[$<R> = 0.99$]{\label{nocoriolis}\includegraphics[height = 7.5cm, width=75 mm]{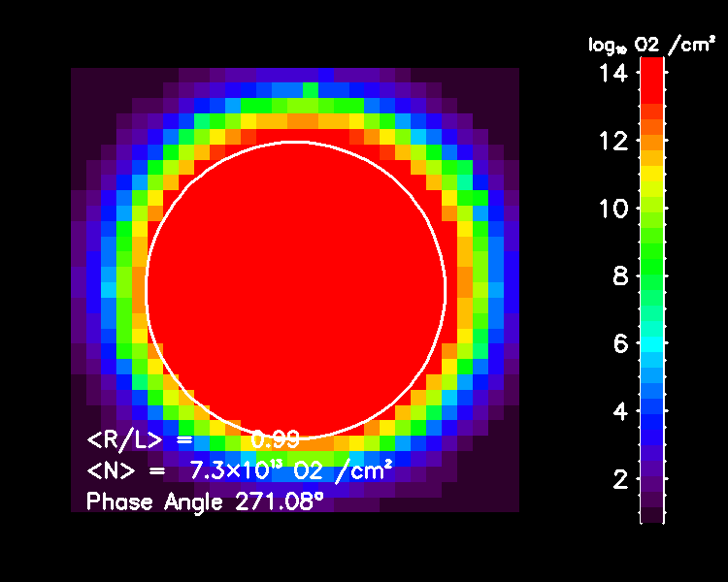}}
\subfigure[$<R> = 6.55$] {\label{coriolis}\includegraphics[height = 75mm, width=75 mm]{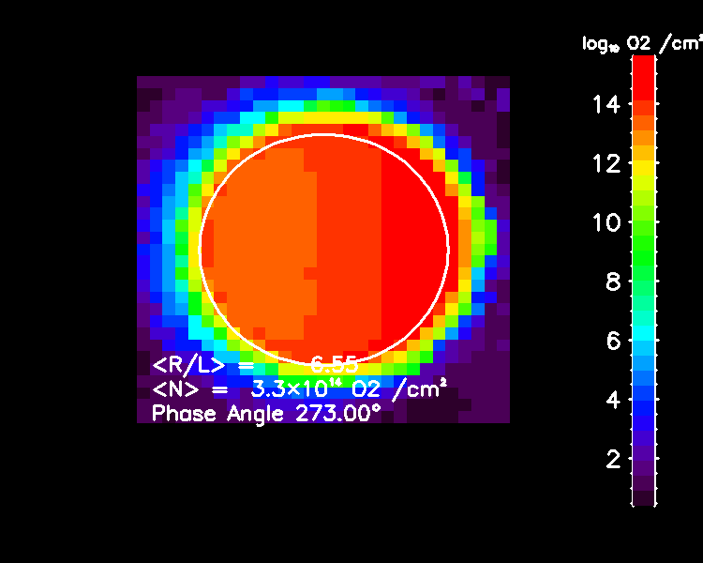}}
\caption{Simulated line of sight column densities of Europa's O$_2$ without (\textbf{a}) and with (\textbf{b}) rotation at the sunlit leading hemisphere $\phi_{obs} = 90^{\circ}$. Both simulations include the non-inertial Coriolis and centripetal forces. The hemispherically averaged dusk-over-dawn column density ratio is indicated in the caption as $<R>$.  }
\end{figure*}

In order to better understand the bulk physical processes regulating the dusk-over-dawn asymmetry we compare hemispherically averaged, dusk-over-dawn ratios for HST observations (black circles) with several EGM simulations assuming isotropic plasma conditions for a range of O$_2$ adsorption conditions (squares, triangles, Xs) in Figure \ref{ddaurora}. The HST data is confined to the near-surface O$_2$ region, and independent of photo-excitation as we uniquely evaluate the forbidden line ratios at 1356 \AA. All simulated data in the shaded region, correspond to a dusk-over-dawn asymmetry, whereas points in the unshaded region represent a dawn enhancement. In our simulated O$_2$ case with the least assumptions (no surface interactions; black squares) the dusk hemisphere is brighter than the dawn hemisphere throughout the \textit{entire} orbit, in the same manner as the HST oxygen observations. Upon varying the initial conditions for our EGM simulations, our results appear to imply that the dusk-over-dawn asymmetry is thermal. That is, the surface temperature regulates the O$_2$ production and the adsorption probability which when varied, strongly influence the diurnal asymmetry. The open squares are identical to the former, nominal case, except that a low temperature-dependence is applied to the O$_2$ yield (q$_{O2}$' = $\frac{q_{O2}}{10}$ in Eqn. \ref{Y}). The reduced thermal dependence is seen to considerably reduce the dusk-over-dawn asymmetry. Conversely, increasing the O$_2$ thermal dependence was found to increase the dusk-over-dawn asymmetry in the 1-D atmospheric evolution model by \citet{Oza18A}. Including the thermal inertia of Europa's ice on the other hand  only increased the asymmetry by $< 7\%$. Furthermore, simply turning on the non-inertial forces in our model (i.e. Coriolis, centripetal) without accounting for Europa's rotation about Jupiter resulted in no dusk-over-dawn asymmetry. The result is shown in Figure \ref{coriolis}.


Next, we evaluate how the dusk-over-dawn asymmetry varies with a variable heat of adsorption for O$_2$.

\subsection{\texorpdfstring{O$_2$ \;} AAdsorption on Grains in the Water Ice Regolith}\label{o2ice}

As discussed in Section \ref{diffusion}, O$_2$ on Europa's $\sim 100 K $ surface is thought to undergo a series of unimpeded ballistic hops due to its low sublimation temperature of $\sim 54 K $. Recently, on the far colder Rhea and Dione, evidence of atmospheric evolution due to O$_2$ adsorption was discussed by \citet{teolis16}. In this regard, we simulate two additional cases where we alter the surface heat of adsorption for the returning O$_2$: $E_{ads} = 0.26 eV$ (Xs) and $E_{ads} (L) \neq E_{ads} (T) $ (triangles). 

The large heat of adsorption case (X) results in stronger interactions with the regolith (i.e. H$_2$O molecules freezing), and in effect longer residence and migration times, which is seen to significantly reduce the dusk-over-dawn asymmetry. The atmospheric bulges described in section 3 then, will not have time to rotate to dusk as the longer timescales will permit destruction near the production region.  
Lastly, we consider a case where the heat of adsorption is non-uniform between the leading and trailing hemispheres (triangles). To test this effect globally we model the heat of adsorption with a simple $cos^2$ function with latitude ($\theta$) and planetary longitude ($\phi^{\prime}$ ): 

\begin{equation}\label{nonU}
E_{ads} (\phi^{\prime}, \theta) = E_1 cos^2\left( \frac{(\phi^{\prime} -\phi_0 ^{\prime})}{2} \right ) cos^2(\theta -\pi/2) + E_0
\end{equation}

Equation \ref{nonU} is a simple surface adsorption relation modeled in the EphiO frame where $\phi_0 ^{\prime} =0$ treats $E_{ads}$ as preferential adsorption on the leading hemisphere and $\phi_0 ^{\prime}=\pi$ as preferential adsorption on the trailing hemisphere. In Fig. \ref{ddaurora} the triangles present a simulation where $E_1 = 0.04eV$ , $E_0 = 0.2 eV$, and $\phi_0 ^{\prime} =0$. It can be seen that the non-uniform heats of adsorption also reduce the dusk-over-dawn asymmetry, and additionally mitigate the \textit{orbital} asymmetries near the leading hemisphere phases. This could imply that the simulated region between $90^{\circ} <\phi_{obs}< 150  ^{\circ}$ possesses stronger O$_2$ interactions with the regolith. Conversely, the trailing hemisphere could also adsorb more efficiently since the column of O$_2$ driven towards leading hemisphere phases is due to the ejection roughly half of a rotation earlier. Thus we are currently unable to deduce whether the O$_2$ adsorbs more efficiently at the leading or trailing surfaces. \\

The transient O$_2$ adsorption simulated here may describe the trapped O$_2$ observed at 5771 \AA as a condensed frost (\citet{spencer02} ; Spencer \& Grundy 2017). That is, the O$_2$ could be trapped in bubbles before it is radiolytically ejected, should the temperature be sufficiently high to allow for vacancies produced by diffusion of the incoming ion \citep{bob97o3}. Recently, Johnson et al. 2018 (submitted) reviewed the origin and fate of the trapped O$_2$ in ice. It was suggested that the observed O$_2$ was trapped at dangling hydrogen bonds, from where it is thermally released.  



\section{Conclusion/Summary}\label{conclusion}

Allowing Europa to \textit{rotate} about Jupiter in our 3-D Monte Carlo simulations of Europa's water product exosphere, results in a dusk-over-dawn O$_2$ asymmetry at all orbital phases. The hemispherically-averaged asymmetry is roughly consistent with the \textit{near-surface} dusk-over-dawn oxygen emission recently observed by HST. We suggest that the near-surface, dusk-over-dawn O$_2$ asymmetry is more generally part of Europa's diurnal cycle of O$_2$, built and destroyed over one orbit. This O$_2$ cycle appears to depend strongly on the (1) O$_2$ production (therefore also the surface temperature), (2) O$_2$ loss, and (3) Europa's rotation rate, all of which lead to a peak in column density at dusk.   \\

The diurnal cycle described in this work, can be summarized as follows : 
\begin{enumerate}
\item \textit{Trailing Day:} The trailing hemisphere source is illuminated and rapidly builds an O$_2$ column rotating with Europa's surface, ejecting between $\sim 2.5 - 3.5 \cdot 10^8$ O$_2$ cm$^{-2}$ s$^{-1}$ throughout the day. 
	\begin{itemize}
    \item After half of an orbit, the rotating O$_2$ column has accumulated an atmospheric bulge of average density $\langle N \rangle_{O_2} \gtrsim 8 \cdot 10^{13}$ O$_2$ cm$^{-2}$ spread across a quarter of a hemisphere from \textit{dusk}, local time. 
	\end{itemize}
\item \textit{Trailing Night:} The trailing hemisphere source enters night, at which point the e$^{-}$-impact dissociation begins to slowly overwhelm the O$_2$ production and acts to isotropically diminish the bulk O$_2$ exosphere. 
\end{enumerate}

A similar dusk-over-dawn asymmetry is predicted at all sub-observer longitudes in Ganymede's near-surface O$_2$ atmosphere \cite{leblanc16}. We note that these O$_2$ bulges maintained near dusk, hinge on the timescales of tidally-locked atmospheres as described by \citet{Oza18A}. The orbital timescale is comparable to the average atmospheric lifetime of the radiolytic O$_2$, so that $\tau_{orb} \sim \tau_{O2}$. In other words, the O$_2$ exosphere survives long enough so that the O$_2$ can accumulate towards dusk during the day, given its simulated thermal dependence.  If the O$_2$ exosphere survived for too long, for instance $\tau_{orb} \ll \tau_{O2}$, the continued production would average out any asymmetries as the O$_2$ loss would be negligible over the orbit.
If the O$_2$ exosphere was destroyed too quickly on the other hand, $\tau_{orb} \gg \tau_{O2}$, the diurnal asymmetries would not have time to build an O$_2$ bulge. Overall, the comparable size of the satellite's rotation rate and the rate at which the near-surface O$_2$ is being lost results in a shift towards dusk. The relative sizes of the rotation rate and the estimated loss rate at Europa and Ganymede, result in a periodic O$_2$ asymmetry, which can be probed in satellite local time.\\

While we have provided evidence for a dusk-over-dawn asymmetry at Europa, a global understanding of Europa's exosphere is far from fulfilled. The O$_2$ cycle depends on the O$_2$ production and loss: source rates may require a largely thermal source (e.g. \citet{Oza18A}, Johnson et al. 2018), and loss rates may need to include plasma interactions (e.g. \citet{dols16}; \citet{l16}). A self-consistent exosphere model including the plasma interaction is therefore sorely needed. A stronger understanding of the spatial morphologies of O$_2$ exospheres will be of particular interest for the spacecraft trajectories at the icy Galilean satellites (i.e. NASA's Europa Clipper; ESA's JUICE).

\textbf{Acknowledgments:}
The authors express their gratitude to Lorenz Roth for insight on better comparing the HST observations and the EGM exosphere simulations, and to Ben Teolis for constructive discussions on O$_2$-water ice interactions.
AVO and FL acknowledge the support of LabEx/ ESEP . This 
work was also supported by CNES "Syst\'{e}me Solaire" 
program. REJ acknowledges support from NASA's Planetary Data Systems Program. This work is also part of HELIOSARES Project supported by the ANR (ANR-09-BLAN-0223) and ANR MARMITE-CNRS (ANR-13-BS05-0012-02). Authors also acknowledge the support of the computational platform 
\texttt{CICLAD} hosted by the Institut Pierre Simon Laplace.

\clearpage

\bibliographystyle{model2-names}\biboptions{authoryear}
\bibliography{bibeuropa_2}

\end{document}